\documentclass[reprint,prl,amsmath,amssymb,aps,showpacs,floatfix]{revtex4-1}
\usepackage{graphicx}
\usepackage{dcolumn}
\usepackage{bm}
\usepackage{natbib}
\usepackage{xcolor}
\usepackage{todonotes}
\begin{document}
\preprint{FFLO}
\title{Calorimetric Measurements of Magnetic-Field-Induced Inhomogeneous Superconductivity Above The Paramagnetic Limit}%

\author{Charles C. Agosta$^1$}
\email{cagosta@clarku.edu}
 \author{Nathanael A. Fortune$^2$}
 \author{Scott T. Hannahs$^3$}
 \author{Shuyao Gu$^2$}
 \author{ Lucy Liang$^2$}
 \author{Ju-Hyun Park$^3$}
 \author{John A. Schleuter$^4$}
\affiliation{$^1$
Physics Department, Clark University, 950 Main Street, Worcester, MA 01610, USA}
\affiliation{$^2$
Physics Department, Smith College, 44 College Lane, Northampton, MA 01063, USA}
\affiliation{$^3$
National High Magnetic Field Laboratory, 1800 East Paul Dirac Drive, Tallahassee, FL 32310, USA}
\affiliation{$^4$ Materials Science Division (MSD-200), Argonne National Laboratory, 9700 South Cass Avenue, Argonne, IL 60439 }


\date{\today}

\begin{abstract}
We report the first magneto-caloric and calorimetric observations of a magnetic-field-induced phase transition within a superconducting state to the long-sought exotic "FFLO"  superconducting state first predicted over 50 years ago. Through the combination of bulk thermodynamic calorimetric and magnetocaloric measurements in the organic superconductor $\kappa$ - (BEDT-TTF)$_2$Cu(NCS)$_2$,  as a function of temperature, magnetic field strength, and magnetic field orientation, we establish for the first time that this field-induced first-order phase transition at the paramagnetic limit $H_p$ for traditional superconductivity is to a higher entropy superconducting 
phase uniquely characteristic of the FFLO state.  We also establish that this high-field superconducting state displays the bulk paramagnetic ordering of spin domains required of the FFLO state.  These results rule out the alternate possibility of spin-density wave (SDW) ordering in the high field superconducting phase.  The phase diagram determined from our measurements --- including the observation of a phase transition into the FFLO phase at $H_p$ --- is in good agreement with recent NMR results and our own earlier tunnel-diode magnetic penetration depth experiments, but is in disagreement with the only previous calorimetric report.   
 
\end{abstract}

\maketitle

Magnetic fields destroy superconductivity. In most cases, this occurs due to the formation of magnetic vortices --- non-superconducting regions containing a magnetic field flux line shielded by circulating electrons --- which increase in density as the magnetic field strength increases, ultimately displacing the superconducting phase.  In the absence of magnetic vortices, the paramagnetic spin susceptibility of the electrons making up the superconducting "Cooper pairs" places another upper limit on superconductivity in magnetic fields. Because the electrons in these pairs have oppositely aligned magnetic moments (spins), the reduction in magnetic energy due to flipping the spin of an individual electron  will exceed the reduction in electronic energy available from the formation of the Cooper pairs above a critical magnetic field $H_P$ known as the Clogston-Chandrasakar paramagnetic limit \cite{clogston62_prl, chandrasekhar1962APL}. A phase transition from the superconducting to the normal metallic state is therefore expected at $H_P$.  

Some 50 years ago, however, Fulde and Ferrell \cite{fulde_ferrell64_pr} and Larkin and Ovchinnikov \cite{LarkinOvchinnikov65JETP} predicted that there might instead be a phase transition at $H_P$ to a different superconducting phase in which  paramagnetic spin domains coexist with a spatially inhomogeneous superconducting phase. This ``FFLO state'' is expected to exist at fields  above $H_P$ in  electronically clean, anisotropic   superconducting materials \cite{BurkhardtRainer1994adp, Matsuda:2007cv, CroitoruHouzet2012prl}.   

The search for inhomogeneous superconductivity has spanned many years, and began in low dimensional single layers of superconducting materials \cite{Tedrow_1973dq}. The first clear calorimetric observations of a bulk field induced phase transition between two superconducting phases were observed in the heavy fermion compound CeCoIn$_5$ \cite{Radovan:2003p67,Bianchi:2003ku}. This transition was initially attributed to  FFLO superconductivity but is now known to correspond to the onset of spin-density wave (SDW) ordering within the superconducting state  \cite{Young:2007cn,  Kenzelmann:2008je, Kenzelmann:2010bq, Tokiwa_2012}. Details of the SDW ordering observed by NMR have led to suggestions of a lower field FFLO transition \cite{Koutroulakis:2010dm} and/or coupling of a FFLO phase along $\vec{H}$ to this SDW transition  \cite{Koutroulakis:2010dm, Kumagai:2011km,Hatakeyama:2015cm}, but there is no clear thermodynamic evidence for these proposals \cite{Tokiwa_2012,Fortune_2014,Kim:2016hy}. The possibility remains of a more complex coupling of the SDW transition to a modified FFLO phase or pair density wave (PDW) \cite{Koutroulakis:2010dm, Kim:2016hy}. More recently, an analog of the FFLO state in the continuously variable parameter space of 
trapped atoms has caught the interest of theorists and experimentalists stimulating a rich new area of research \cite{PhysRevLett.104.236402, LiaoRittner2010Nature, OlsenRevelle2015PRA}.

 \begin{figure} 
\centering
\includegraphics[width=0.45\textwidth]{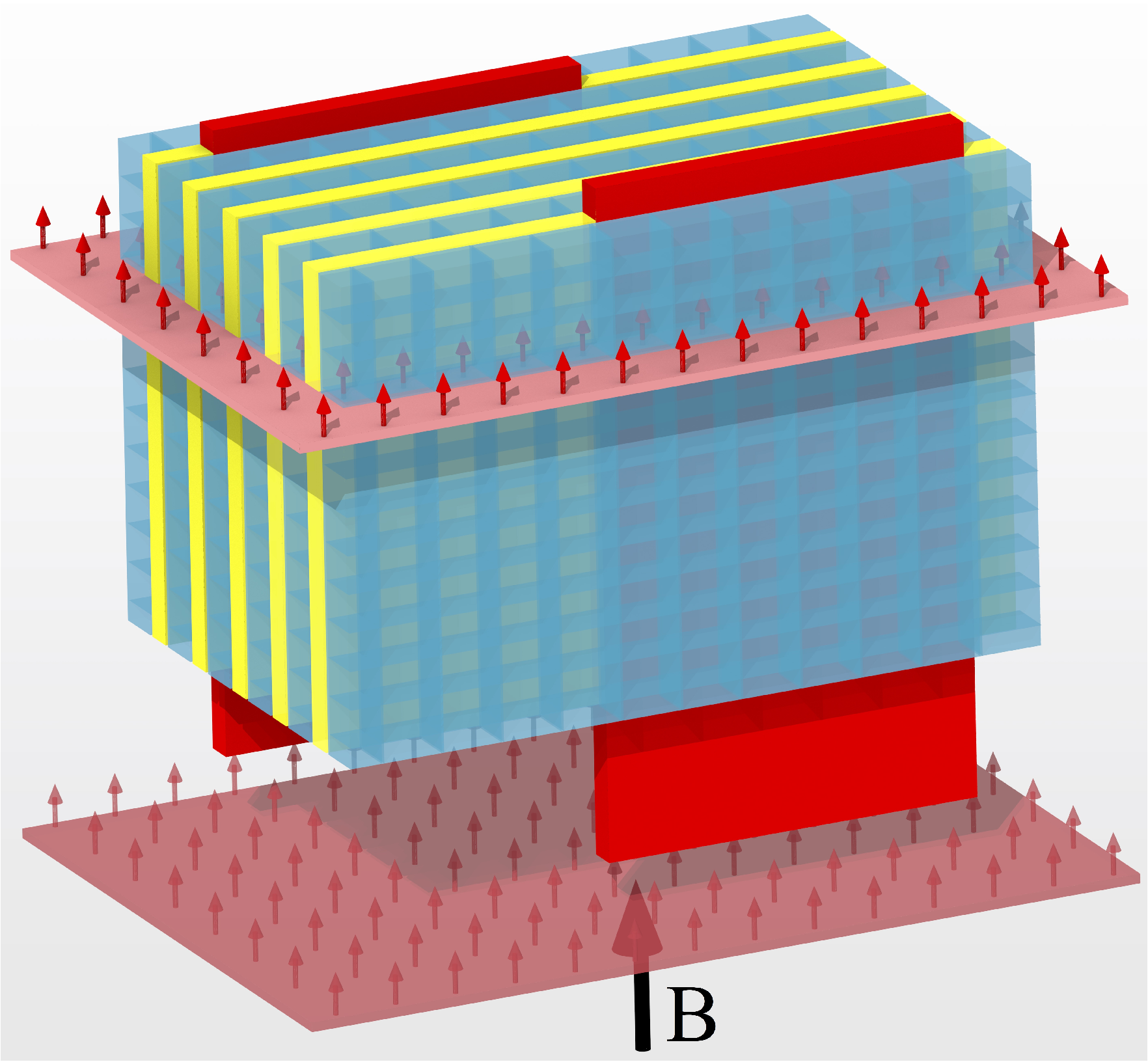}
\caption{\label{fig:ffloCartoon} Color online. Cartoon of the FFLO state showing the nodes in the order parameter as horizontal planes where we estimate the spin-polarization to be $\approx 10\%$ at 25 T in the low temperature limit. The red arrows represent the net spin polarization. 
Although the diagram is schematic, all of the lengths are to scale; small boxes represent the unit cells of $\kappa$ - (BEDT-TTF)$_2$Cu(NCS)$_2$, yellow slabs represent the least conducting layers of the crystal, and  red rectangles represent Josephson core-less vortices at about the right distance apart in a 25 T field. Full height of the crystal is $\approx \textrm{\ 20 nm}$. }
\end{figure} 

A cartoon of a FFLO state in a quasi-2D superconductor is shown in Fig. \ref{fig:ffloCartoon} for the simplest possible case: one in which the spatial modulation of the order parameter occurs in the direction of the applied magnetic field. Paramagnetic spin domains form in normal state regions existing at the nodes in the superconducting order parameter, while sheet-like coreless Josephson vortices form between the 2D superconducting layers. At least two criteria must be satisfied to create this in-homogeneous superconducting state. The first is the suppression or exclusion of vortices, as vortices normally destroy superconductivity at magnetic fields lower than the $H_P$ limit. The second criteria is that the materials need to be sufficiently electronically clean for a coherent superconducting wave function to persist over distances on the order of the FFLO wave vector (corresponding to the distance between the nodes in Fig. \ref{fig:ffloCartoon}).

Quasi-2D layered organic superconductors should therefore be perfect candidates for forming a FFLO state: they have long electronic mean free paths, as shown by large quantum oscillations \cite{Singleton:2002p45} and they are highly two-dimensional, so the vortices can be confined to the least conducting layers \cite{Agosta_2012}. As a consequence, the vortices become Josephson coupled \cite{kirtley_moler1999} and only weakly interact with the superconducting layers \cite{mansky_chaikin93_prl}.  Phase diagrams suggesting the existence of a higher field superconducting phase in $\kappa$ - (BEDT-TTF)$_2$Cu(NCS)$_2$ ($T_c$ of 9.5 K) and other superconductors have been established using rf penetration depth\cite{Singleton_2000}, tunnel diode oscillator (TDO) rf penetration depth \cite{Cho_2009, martin_agosta05_prb, Coniglio_2011, Agosta_2012}, resistivity\cite{Tsuchiya_2015hb}, thermal conductivity \cite{Tanatar:2002bn}, heat capacity \cite{Lortz_2007}, torque magnetometry \cite{Bergk_2011, Tsuchiya_2015hb}, and NMR \cite{Wright_2011,Mayaffre_2014}, but the claimed location, slope, and curvature of the phase boundary between the two superconducting states varies dramatically. 

In this paper, we report results of magnetic-field-dependent heat capacity and magnetocaloric effect measurements as a function of field strength, direction, and temperature between 0.15 K and 4.2 K. 
The heat capacity measurements allow us to discern the locations of the phase boundaries and determine the order of the transitions. The magnetocaloric effect measurements allow us to infer whether a phase is paramagnetic, diamagnetic,  or fully polarized, and, for first order transitions, observe the sign of the change in entropy at the transition.   

To carry out these measurements, we have made use of a recently developed rotatable calorimeter \cite{Fortune_2014} designed for use in the portable dilution refrigerator and 32 mm bore high field resistive magnets at the National High Magnetic Field Laboratory dc field facility. The calorimeter fits into a top-loading single-axis probe  \cite{Palm_1999} capable of 360 degree rotation at base temperature with a resolution of 0.02 degrees.  The sample is weakly thermally linked to a temperature controlled platform inside the vacuum calorimeter, which is in turn weakly linked to the cryogenic mixture. When inserted into a dilution refrigerator, measurements can be made from 100 mK to 10 K during a single experiment.  The heat capacity was measured as a function of magnetic field for a series of fixed temperatures and field orientations using an ac calorimetric method \cite{Sullivan_1968}, after corrections  for the magnetic-field dependence of the resistive thermometers \cite{Fortune_2000}. The sample layers were oriented parallel to the applied magnetic field to within $0.1$ degrees through the calorimetric determination of $H_{c2}(\theta)$.

\begin{figure} 
\centering
\includegraphics[width=0.45\textwidth]{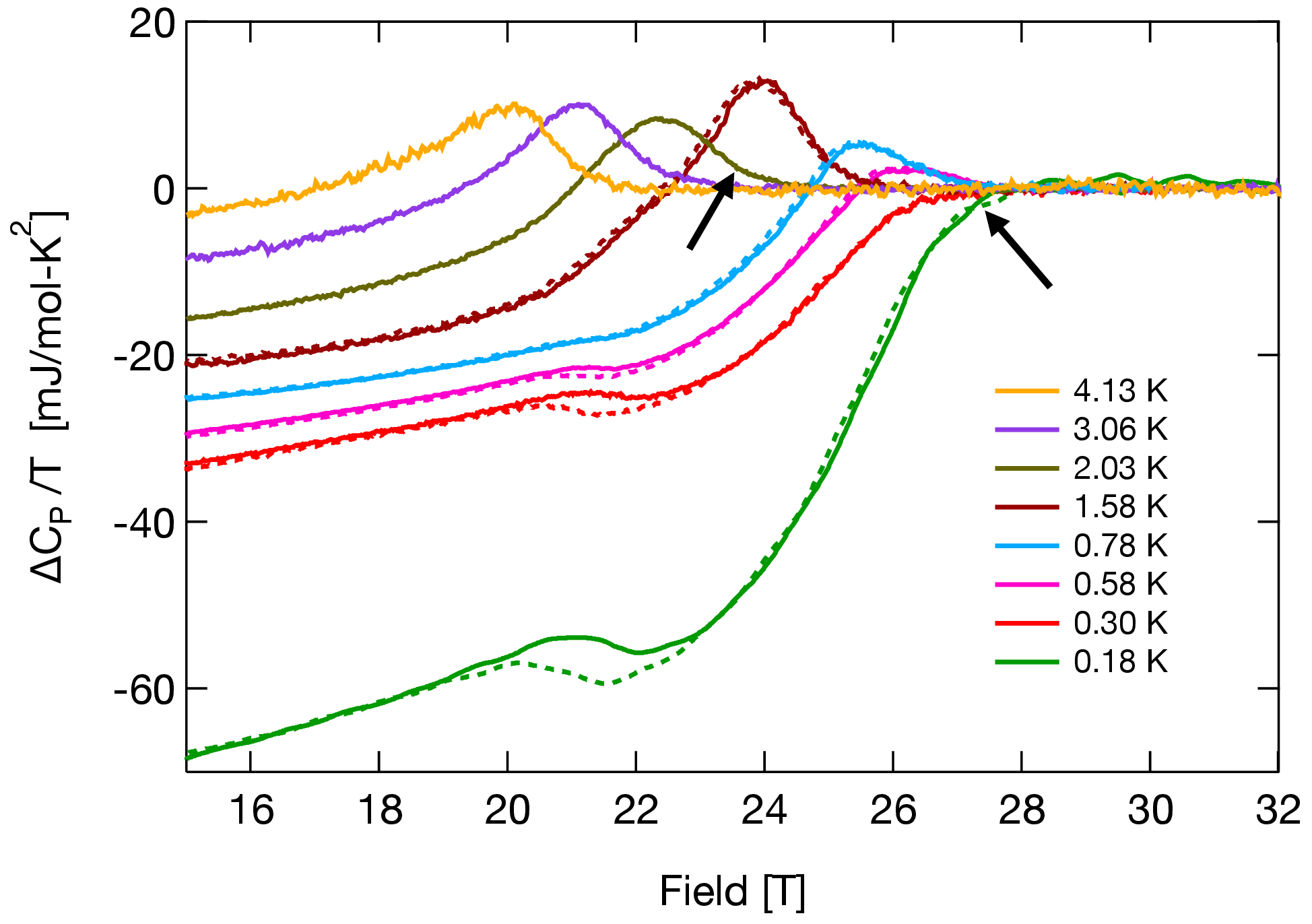}
\caption{\label{fig:CpVSfld} Color online. Magnetic-field induced change (solid-up, dotted-down) in the specific heat of $\kappa\textrm{-(BEDT-TTF)}_2\textrm{Cu}{\textrm{(NCS)}_2}$ (scaled by temperature) for magnetic field parallel to the superconducting planes ($\theta = 0$). For ease of comparison, $\Delta C_p /T$ is set equal to zero  in the normal, non-superconducting state ($ H > H_{\textrm{c2}}$). We observe a first order phase transition between two different superconducting states at $H  \approx H_p = 20.7 \textrm{ T}$, followed by a  transition to the normal state at a temperature dependent field $H_{c2}(T)$. Arrows represent $H_{c2}$ for 0.18 K and 2.03 K.} 
\end{figure}  

\begin{figure}  
\centering
\includegraphics[width=0.45\textwidth]{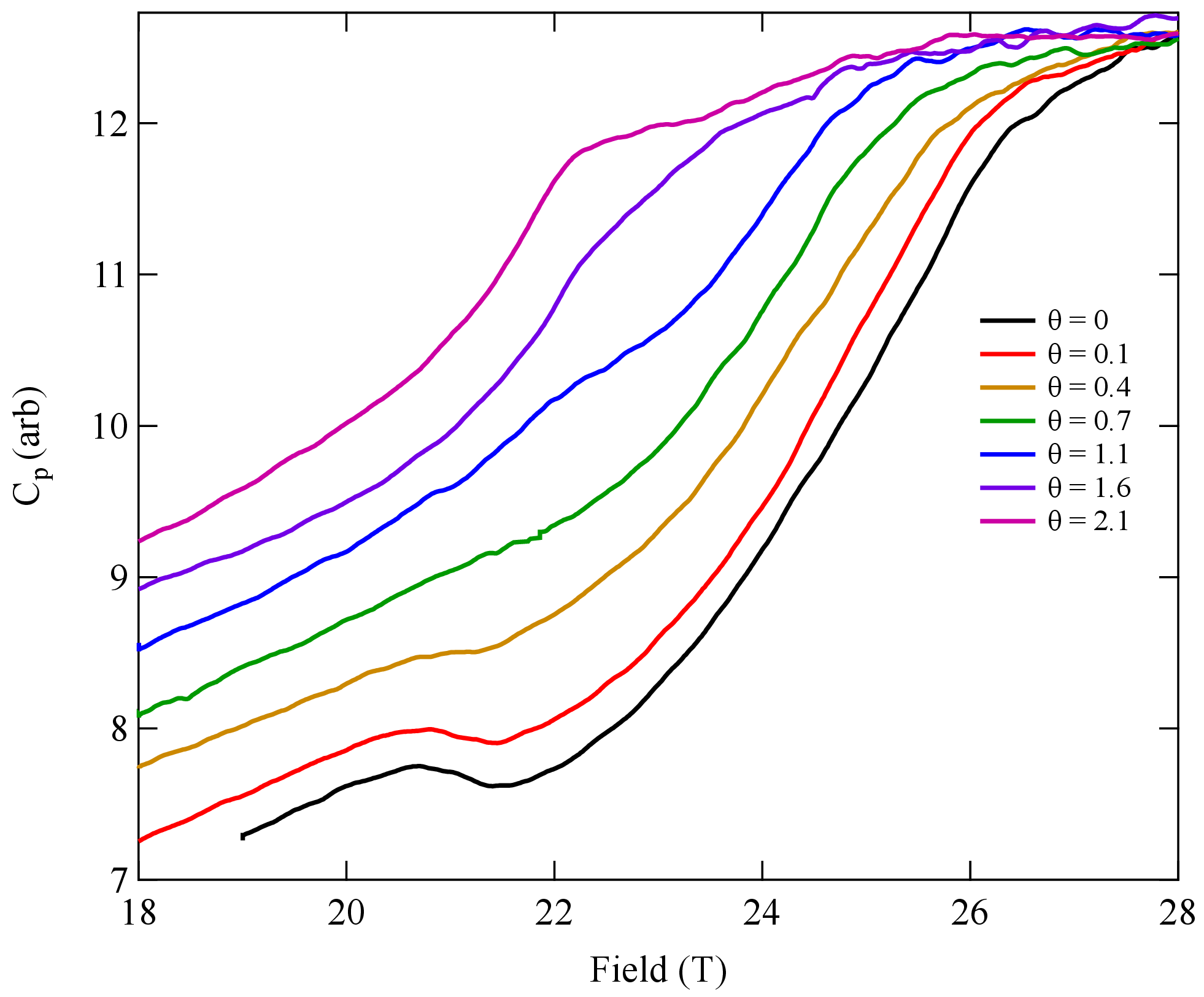}
\caption{\label{fig:angleDep} Color online. Evolution of the heat capacity sweeps as a function of angle at 0.3 K, with plane parallel $\theta = 0$ orientation shown in black at the bottom. The bump at 21 T marking the transition at $H_p$ at $0^{\circ}$ is gone by $0.7^{\circ}$ and is replaced by the beginning of the $H_{c2}$ transition by $1.6^{\circ}$, consistent with expectation for 2D FFLO superconductivity.}
\end{figure}

Figure \ref{fig:CpVSfld} shows the measured magnetic-field-dependent heat capacity between 15 and 32 T in the low temperature limit for an applied field parallel to the conducting layers. The results reveal the presence of a hysteretic, and therefore first order  phase transition between two different superconducting states at $H  \approx H_p = 20.7 \textrm{ T}$ followed by a phase transition to the normal, non-superconducting state at a temperature dependent field $H_{c2}(T)$.   In addition, we note that this transition to a high field superconducting state is strongly angle dependent, disappearing for angles $\theta \geq 1^{\circ}$ as shown in Fig. \ref{fig:angleDep}. The rapid disappearance of the transition at $H_P$ as the sample is rotated away from the field parallel orientation is as expected for an FFLO transition in a 2D superconductor, since such a phase would readily be destroyed by spin-orbit scattering once the sample is tilted enough for Abrikosov vortices to begin to penetrate the superconducting planes. 

The overall increase in specific heat with increasing applied field is expected because the magnetic field is breaking Cooper pairs (and when in the FFLO phase, creating paramagnetic spin domains), thereby increasing the number of quasiparticles that can carry entropy. The steep upward curvature to the field dependence of $C_p(H)$ that arises at high fields within the superconducting state is characteristic of 
strongly Pauli-paramagnetic superconducting materials \cite{Ichioka:2007kl, Machida:2008cy}. At higher temperatures we observe a broad peak in the specific heat $C_p(H){\vert}_T$ due to the transition from the high-field superconducting to normal state at $H_{c2}(T)$. At lower temperatures, however, the peak diminishes in height, disappearing by 0.3 K. For consistency, we therefore take $H_{c2}(T)$ to correspond to the inflection point between the normal and superconducting state, as shown by arrows in Fig.~\ref{fig:CpVSfld}. 

\begin{figure} 
\centering
\includegraphics[width=0.45\textwidth]{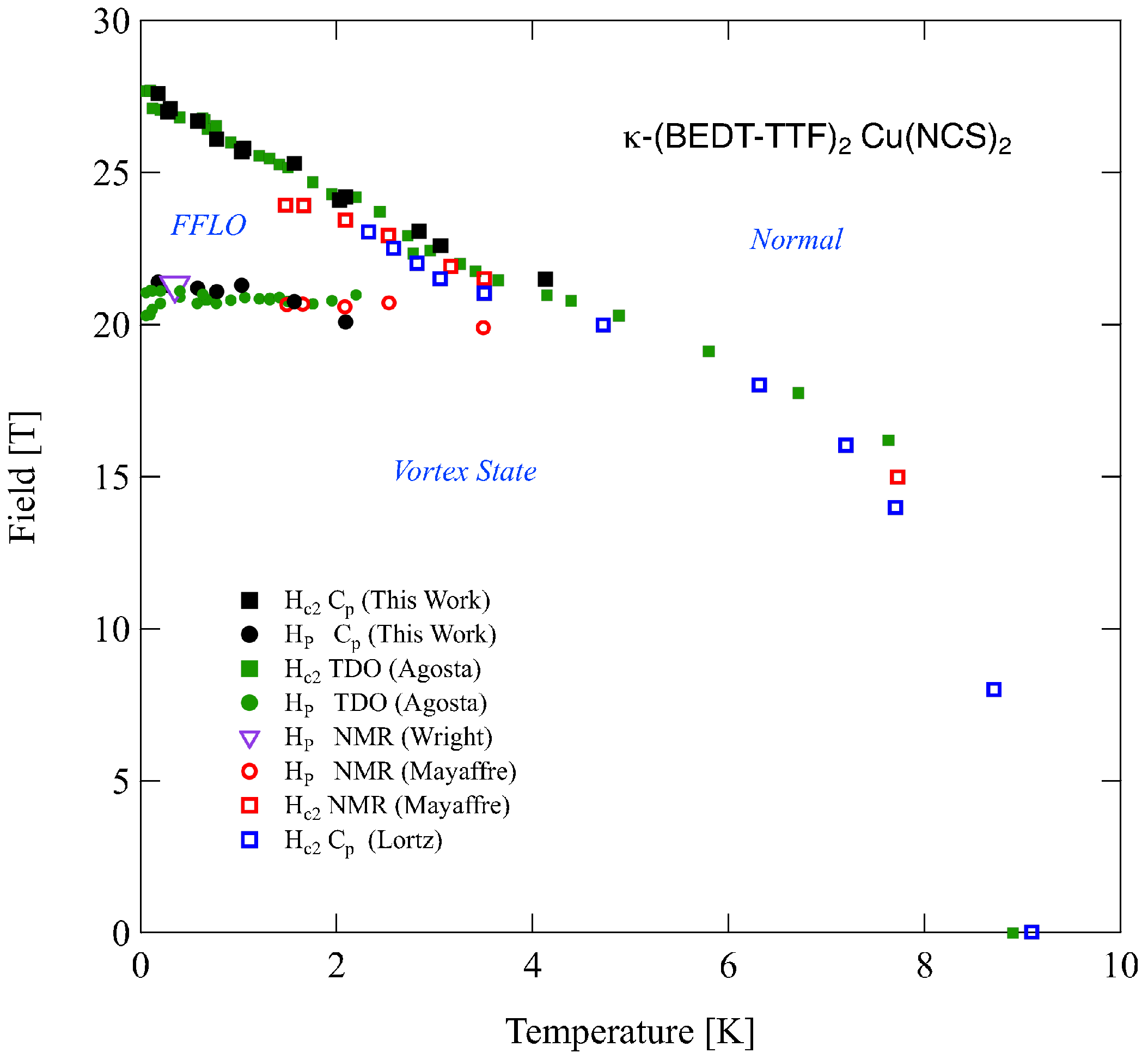} 
\caption{\label{fig:phaseDiagram} Color online. Phase diagram of $\kappa$-(BEDT-TTF)$_2$Cu(NCS)$_2$ for parallel magnetic field ($\theta = 0$). Solid black circles represent our calorimetric observations of the phase transitions between the lower and higher field superconducting phases at $H_p$, and squares, the normal and superconducting state at  $H_{c2}(T)$. Points from an earlier calorimetric determination of $H_{c2}(T)$ \cite{Lortz_2007} are shown as open blue squares. Also included are supporting determinations of both the $H_{c2}$ and $H_p$ phase boundaries by means of rf penetration measurements (green) \cite{Agosta_2012} and NMR measurements \cite{Wright_2011,Mayaffre_2014} (open purple and red symbols, respectively).}
\end{figure}

The superconducting phase diagram thus determined from our calorimetric measurements is shown in Fig. \ref{fig:phaseDiagram}. For comparison, we also include data points from earlier NMR \cite{Mayaffre_2014,Wright_2011}, rf penetration depth \cite{Agosta_2012}, and specific heat measurements \cite{Lortz_2007}. Our $H_{c2}(T)$ phase boundary is in agreement with previous measurements. The location and curvature of our phase boundary between the low-field superconducting states and suggested FFLO state at $H_P$ is in good experimental agreement with our earlier tunnel diode oscillator (TDO) rf penetration measurements \cite{Agosta_2012} and NMR  \cite{Wright_2011,Mayaffre_2014} but in strong disagreement with previous penetration depth \cite{Singleton_2000} and calorimetric \cite{Lortz_2007} claims for observation of an FFLO phase boundary.    

The size of the specific heat jump $\Delta C/T$  we observe at $H_{c2}$ is governed by the magnetic Ehrenfest relation \cite{Kumar:2009fh} for a second order phase boundary: 
\begin{equation}
\left({\frac{\Delta C}{T}}\right) = \left(\frac{\partial \Delta M}{\partial H} \right) \left(\frac{\partial H_{c2}}{\partial T}\right)^2. 
\end{equation}

For traditional superconductors, the first term in this product remains finite, but the second term --- the slope of the $H_{c2}(T)$ phase boundary --- starts high, then flattens out as $T \rightarrow 0$ \cite{werthamer_helfand66_pr}, leading to a decrease in $\Delta C/T$ as $T \rightarrow 0$.  
In contrast, in the FFLO state, the slope of $H_{c2}(T)$ remains high \cite{Shimahara:1994hk}, while the first term in the product approaches zero as $T \rightarrow 0$ \cite{Gruenberg:1966cg, Wright_2011}, leading to the observed change in the field position of the specific heat jump and the absence of a discernible jump by 0.3 K.

When we transform  our field sweep measurements at fixed T into temperature sweeps at fixed field, we find that 
in the region of overlap ($T \geq 1.8 \textrm{ K}$), the field dependence of our data is self consistent with an earlier report by Lortz \textit{et.~al} \cite{Lortz_2007}, including the broad peak below  $H_{c2}$. The  weak temperature dependence of the first order phase line we observe at $H_P$ means however, that this transition would not be expected to be resolved in their temperature sweeps of $C_p(T)$ at constant magnetic field\cite{Lortz_2007}.  In contrast, their temperature sweep measurements naturally yield sharper peaks at $H_{c2}(T)$ than field sweeps as $T \rightarrow T_c$. 

We now turn to our swept-field magnetocaloric measurements across this low field to high field superconducting phase boundary.  Magnetocaloric measurements were made in the field parallel $\theta = 0$ orientation. In these measurements, as with the specific heat measurements, the sample is thermally linked to a temperature controlled platform while the magnetic field is swept up or down, but in contrast to the specific heat measurements, no  heating is provided by the sample heater. The measured temperature difference $\Delta T$ between the sample and the platform depends on the field sweep rate $\dot{H}$, the thermal conductance $\kappa$ of the wires linking the sample and platform, and the temperature dependence of the magnetization $\left(\partial M/\partial T\right)_H$ \cite{Fortune:2009p365}:

\begin{equation}
\label{MCE_equation}
\Delta T = - \left[\frac{T}{\kappa}\left(\frac{\partial M}{\partial T}\right)_H + \tau\frac{d\Delta T}{dH}\right]\dot{H}
\end{equation}
where $\tau = C/\kappa$ is the sample to platform relaxation time. 

For a strongly temperature-dependent paramagnetic phase ($\partial M/\partial T < 0$) and sufficiently high sweep rate, the up sweep will therefore be warmer than the down sweep. At a first order transition, additional contributions to the magnetocaloric effect arise from (1) the release of latent heat at a first-order transition upon leaving a higher-entropy phase, (2) the absorption of latent heat upon entering a higher-entropy phase,  and (3) the release of heat in both sweep directions due to irreversibility, reflecting the system's tendency to briefly remain 
at the boundary in what becomes a metastable state before jumping to the lower-energy thermodynamically preferred state. 

The expected change in entropy at the transition depends on the nature of the high field phase. 
In the FFLO state, the SC gap function is inhomogeneous (being spatially modulated with a wave length $2\pi/q$); paramagnetic quasiparticles appear periodically at the nodes in the gap function \cite{Tokiwa_2012}. These additional quasiparticles lead to an increase in entropy upon crossing the phase boundary at $H_p$ for up sweeps into a FFLO phase, and a corresponding decrease in entropy at $H_p$ for down sweeps \cite{Cai_2011,Tokiwa_2012}. In contrast, spin-density wave ordering within a homogeneous superconducting phase leads to a reduction in the number of degrees of freedom and a corresponding decrease in entropy for the high field phase \cite{Tokiwa_2012}. 

\begin{figure}  
\centering
\includegraphics[width=0.45\textwidth]{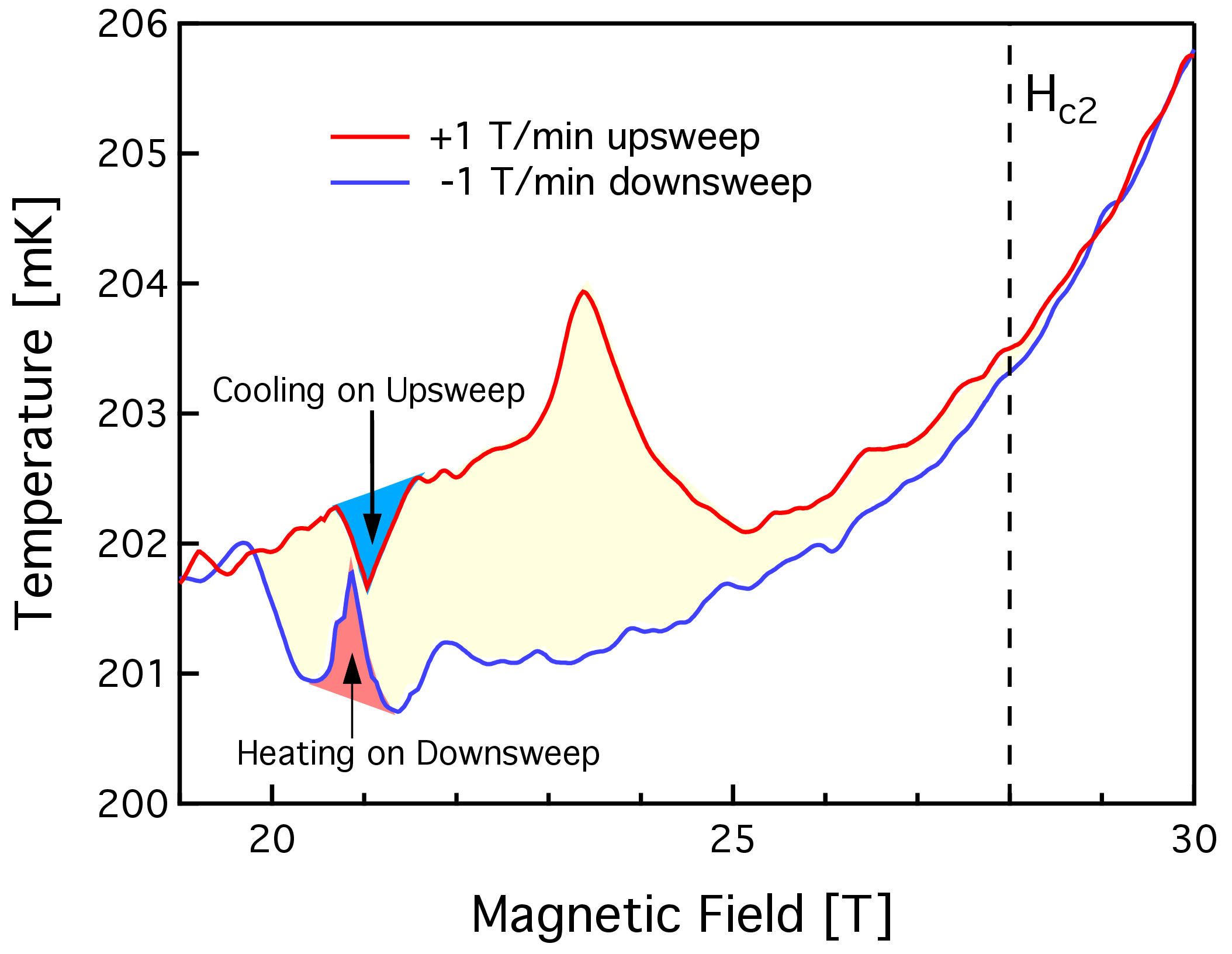}
\caption{\label{fig:MCE} Color online. Magnetocaloric effect measurements at 200 mK. The  temperature difference $\Delta T$ between the up and down sweeps indicates that the system has become paramagnetic (yellow shading). On down sweeps (blue line) there is a brief \textit{increase} in the temperature of the sample  due to \textit{release} of latent heat from the sample at $H_p$. On up sweeps (red line) there is a corresponding \textit{decrease} in the temperature of the sample due to \textit{absorption} of latent heat by the sample at $H_p$. The results indicate the high field state is higher entropy than the low field state, as expected for high field FFLO superconductivity. }
\end{figure}

As seen in Fig.~\ref{fig:MCE}, we find a positive difference in sample temperature $\Delta T_{ud} = \Delta T_{up} - \Delta T_{down}$ emerging as $H \rightarrow H_P$, consistent with the emergence of  paramagnetic spins. The strong maximum in $\Delta T_{ud}$ within the paramagnetic high field state is possibly due to the remarkably strong temperature dependence of the electronic spin polarization (and spin relaxation rate) observed in NMR at these fields \cite{Wright_2011,Mayaffre_2014}. At a still higher field, $\Delta T_{ud} \rightarrow 0$ as $H \rightarrow H_{c2}$ since $\partial M/\partial T \rightarrow 0$ due to the strong polarization of the high field, low temperature metallic state. Superimposed on that overall positive temperature difference $\Delta T_{ud}$ we also observe the release of latent heat (plus irreversibility heating) at the $H_p$ boundary on the down sweep its absorption (less irreversibility heating) on the up sweep, as expected for the high field FFLO state.
Measurements on a second sample gave the same results. 

This directly observed increase in entropy upon entering the high field superconducting state implies, by the magnetic Clausius-Clapeyron equation, that the phase boundary between the two superconducting states must be at least weakly negatively temperature dependent. This result is in agreement with the FFLO phase diagram presented here in addition to theoretical\cite{werthamer_helfand66_pr, Tokiwa_2012} and  experimental\cite{CoffeyMartin2010} expectation. 

We have shown (1) that a bulk thermodynamic first-order phase transition occurs within the superconducting state of the molecular superconductor $\kappa$-(BEDT-TTF)$_2$Cu(NCS)$_2$ (2) that this transition occurs at the paramagnetic limit $H_p$ for traditional superconductivity in this material (3) that this phase transition occurs only when vortices are excluded from the 2D superconducting planes (4) that this high-field superconducting state is paramagnetic, and (5) that this high-field superconducting state is higher entropy,  even though higher magnetic fields usually decrease entropy. 
Taken together, these results provide the first thermodynamic case for the existence of an inhomogeneous FFLO superconducting phase with paramagnetic spin domains for $H \geq H_p$ in highly-anisotropic 2D molecular superconductors, such as $\kappa$ - (BEDT-TTF)$_2$Cu(NCS)$_2$.

\section{acknowledgments}The authors thank S. Blundell for helpful discussions. A portion of this work was performed at the National High Magnetic Field Laboratory, which is supported by National Science Foundation Cooperative Agreement No. DMR-11157490 and the State of Florida.

\bibliography{FFLO_bibliography}
\bibliographystyle{apsrev}

\end{document}